\documentclass[aps,prl,twocolumn,superscriptaddress, showpacs]{revtex4}

\usepackage{graphicx}
\usepackage{color}
\usepackage{color}
\DeclareGraphicsExtensions{.jpg,.pdf,.png,.eps}

\begin{document}

\title{Searching for thermal signatures of persistent currents in normal metal rings}

\author{Germain M. Souche}
 \affiliation{Institut N\'EEL, CNRS-UJF, 25 rue des Martyrs, BP166, 38042 Grenoble, France.}
 \author{Julien Huillery}
 \affiliation{Universit\'e de Lyon, Ecole Centrale de Lyon, Laboratoire Amp\`ere, 36 Avenue Guy de Collongue, 69134 Ecully, France.}
 \author{H. Pothier}
\affiliation{Quantronics Group, Service de Physique de l'État Condensé (CNRS, URA 2464), IRAMIS, CEA-Saclay, 91191 Gif-sur-Yvette, France}
\author{Philippe Gandit}
\affiliation{Institut N\'EEL, CNRS-UJF, 25 rue des Martyrs, BP166, 38042 Grenoble, France.}
\author{J\'{e}r\^ome I. Mars}
\affiliation{GipsaLab, Grenoble INP and Universit\'{e} Joseph Fourier, CNRS UMR 5216, Domaine Universitaire, 38402 Saint Martin d'H\`eres, France}
\author{Sergey E. Skipetrov}
\affiliation{Universit\'e Grenoble 1/CNRS, LPMMC UMR 5493 - B.P. 166, 38042 Grenoble, France}
\author{Olivier Bourgeois} \email{olivier.bourgeois@grenoble.cnrs.fr}
\affiliation{Institut N\'EEL, CNRS-UJF, 25 rue des Martyrs, BP166, 38042 Grenoble, France.}

\date{\today}

%
\pacs{ 73.23.-b, 73.23.Ra, 68.65.-k, 64.70.Nd}

\begin{abstract}
We introduce a calorimetric approach to probe persistent currents in normal metal rings. The heat capacity of a large ensemble of silver rings is measured by nanocalorimetry under a varying magnetic field at different temperatures (60~mK, 100~mK and 150~mK). Periodic oscillations versus magnetic field are detected in the phase signal of the temperature oscillations, though not in the amplitude (both of them directly linked to the heat capacity). The period of these oscillations ($\Phi_0/2$, with $\Phi_0 = h/e$ the magnetic flux quantum) and their evolution with temperature are in agreement with theoretical predictions. In contrast, the amplitude of the corresponding heat capacity oscillations (several $k_{\mathrm{B}}$) is two orders of magnitude larger than predicted by theory.
\end{abstract}

\maketitle

\section{Introduction}

At very low temperatures $T$, a small isolated metal ring carries a weak but non-dissipative current that is not destroyed even by the presence of disorder in the ring \cite{Butti83}. It follows from very general arguments that this persistent current $I$ is periodic in the magnetic flux $\Phi$ threading the ring \cite{Imry02,Akker07}:
\begin{eqnarray}
I(\Phi, T) = \sum\limits_{m=1}^{\infty} I_m(T) \sin\left(2 \pi m \frac{\Phi}{\Phi_0} \right),
\label{pcperiod}
\end{eqnarray}
where $\Phi_0 = h/e$ is the magnetic flux quantum. The typical magnitude of the persistent current in a ring of circumference $L$ is $I \sim e E_{\mathrm{Th}}/\hbar$, where $E_{\mathrm{Th}} = \hbar D/L^2$ is the Thouless energy and $D$ is the diffusion coefficient determined by the scattering of electrons from the disorder in the ring. For $L \sim 1$ $\mu$m, we find $I \sim 1$ nA. The precise value of $I$ and the direction in which this current is flowing are determined by the configuration of impurities and are thus random.

Due to the weak magnitude of the persistent current, only the first two harmonics $m = 1$, 2 in Eq.\ (\ref{pcperiod}) were observed experimentally \cite{Levy90,Chandra91,benoit93,Jariwala01,Deblock02,Bluhm09,Bleszynski09}. They were also in the focus of numerous theoretical studies (see Refs.\ \cite{Cheung89,Ambe90,Ambe90epl,Eckern91,Riedel93,motta,Bary08,Ginossar10} for a representative selection of theoretical works). The ensemble of the existing literature indicates that the statistical properties of the first harmonics $I_1$ can be understood within the model of noninteracting electrons, whereas the second harmonics $I_2$ is dominated by interaction effects. In the experiments performed on single rings \cite{Chandra91,Bluhm09} or on small ensembles of identical rings \cite{Jariwala01,Bleszynski09}, both the first and the second harmonics of $I$ have been detected. Hence, the measured period of current oscillation with $\Phi$ is $\Phi_0$. The mean values of $I_1$ and $I_2$, their variances, and possibly the full probability distributions can be measured \cite{Ginossar10}. On the contrary, if large ensembles of rings (up to $10^7$ rings in Ref.\ \cite{Levy90}) are used \cite{Levy90,Deblock02}, averaging over all the rings of the ensemble inevitably takes place and one only has access to the average current $\langle I \rangle$. Because $\langle I_1 \rangle = 0$ \cite{footnote}, the measured signal is mainly due to the second harmonics $\langle I_2 \rangle \ne 0$. The observed current then oscillates with a period $\Phi_0/2$.

The properties of the first harmonics $I_1$ of the persistent current are nowadays relatively well understood. If early observations were somewhat contradictory \cite{Chandra91,Jariwala01}, more recent works have demonstrated an impressive agreement between experiment and the noninteracting electron theory \cite{Bluhm09,Bleszynski09,Ginossar10}. The second harmonics $I_2$, on the contrary, still represents a challenge for both theory and experiment. Even though a consensus exists on the importance of electron-electron interactions to explain its properties, the calculation assuming {\em repulsive\/} interactions \cite{Ambe90,Eckern91} yields the magnitude of $\langle I_2 \rangle$ which is significantly smaller than the one measured in the experiments. In addition, the paramagnetic response of the rings at small magnetic field predicted by the theory, disagrees with observations \cite{Levy90,Deblock02}. Diamagnetic response can be obtained in a theory assuming {\em attractive\/} interactions \cite{Ambe90epl}. However, interactions which are sufficiently strong to reproduce the experimentally observed values of $\langle I_2 \rangle$ would induce a transition to the superconducting state at temperatures which are too high to be compatible with known properties of (some of the) metals used in the persistent current experiments: copper, gold, and silver. Indeed these metals do not exhibit superconductivity even at the lowest temperatures accessible experimentally (down to 0.1 mK \cite{motta}). A possible solution to this problem has been recently proposed by Bary-Soroker {\em et al.} \cite{Bary08}: a tiny amount of magnetic impurities can destroy the superconductivity but has little impact on the persistent current. Two new parameters --- the spin-scattering rate $1/\tau_s$ and the bare superconducting transition temperature $T_c^0$ of the material without magnetic impurities --- appear in the theory and allow for a reasonable explanation of the experiments reported in Refs.\ \cite{Levy90} and \cite{Jariwala01}.

Motivated by the recent progress in the research on persistent currents in normal metal rings, we propose here a new way of detecting these currents. Our approach is radically different from the one employed in all previous experiments: if these anterior works relied on the measurement of the rings' magnetic moment (either using a more or less sophisticated version of a SQUID magnetometer \cite{Levy90,Chandra91,Jariwala01,Bluhm09}, coupling the rings to a superconducting microresonator \cite{Deblock02}, or using an elegant micromechanical detector \cite{Bleszynski09}), we propose to focus on the rings' heat capacity. Our idea stems from the basic principles of thermodynamics. On the one hand, the heat capacity at constant pressure is given by $C_p = -T (\partial^2 F/\partial T^2)_p$, where $F$ is the thermodynamic free energy. On the other hand, the persistent current is $I = -\partial F/\partial \Phi$. Therefore, we find
\begin{eqnarray}
\frac{\partial C}{\partial \Phi} = T \frac{\partial^2 I}{\partial T^2}.
\label{maxwell}
\end{eqnarray}
From here on we omit the subscript `p' of the heat capacity to lighten the notation. 
Equations (\ref{pcperiod}) and (\ref{maxwell}) imply that
\begin{eqnarray}
C(\Phi, T) &=& C(0, T)
\nonumber \\
&+& \sum\limits_{m=1}^{\infty} C_m(T) \left[ 1 - \cos\left(2 \pi m \frac{\Phi}{\Phi_0} \right) \right],
\label{cpperiod}
\end{eqnarray}
where $C_m(T) = (T \Phi_0/2\pi m ) \partial^2 I_m(T)/\partial T^2$. The heat capacity $C$ is therefore also a periodic function of the magnetic flux $\Phi$.

Equation (\ref{maxwell}) shows that the dependence of $C$ on $\Phi$ is intimately related to the dependence of $I$ on $T$. This link between $C$ and $I$ was already exploited by Yang and Zhou to investigate the impact of spin-orbit coupling on both the persistent current and the heat capacity in the model of noninteracting electrons \cite{Yang96}, as well as by (some of the) present authors to study the entrance of magnetic vortices into superconducting loops \cite{Bourg05,Ong06,Ong07}. The experiments reported here are, however, much more involved than those performed in the superconducting state: not only we work at much lower temperatures ($T \sim 100$ mK instead of $T \sim 1$ K), but also the variations of the heat capacity to detect are much weaker ($\Delta C \sim k_B$ instead of $\Delta C \sim 10^3$ $k_B$).

\begin{figure}
\begin{center}
 \includegraphics[width=\columnwidth]{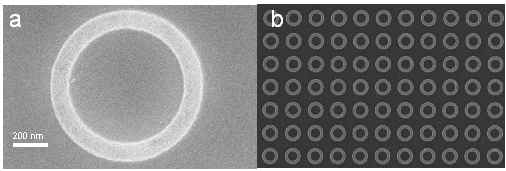}
 \end{center}
 \caption{(a) Scanning electron microscope (SEM) image of a single silver mesoscopic ring, the scale bar represents 200~nm. (b) SEM image of the array of rings.}
 \label{anno}
\end{figure}

\section{Experimental setup}

Our sample (Fig.~\ref{anno}) is composed of $N = 5 \times 10^6$ noninteracting silver rings ($2 r = 600$ nm diameter, $w \simeq 140$~nm arm width, $d = 34$ nm thickness, total mass $m = 330$~ng). The silver rings are deposited by e-gun evaporation under $10^{-6}$~mb vacuum. Thanks to the high purity of the silver the phase coherence length is large, of about 10 $\mu$m at the temperature of 100~mK, which is far above the diffusive ring circumference $L$ \cite{lphi}. The specific heat of silver at $T = 0.1$~K can be estimated to be of the order of $10^{-6}$~J/gK, and hence the magnetic-field-independent part of the heat capacity of the rings is expected to be  $3.3 \times 10^{-13}$~J/K; representing approximately 10\% of the total heat capacity.
\begin{figure}
\begin{center}
 \includegraphics[width=\columnwidth]{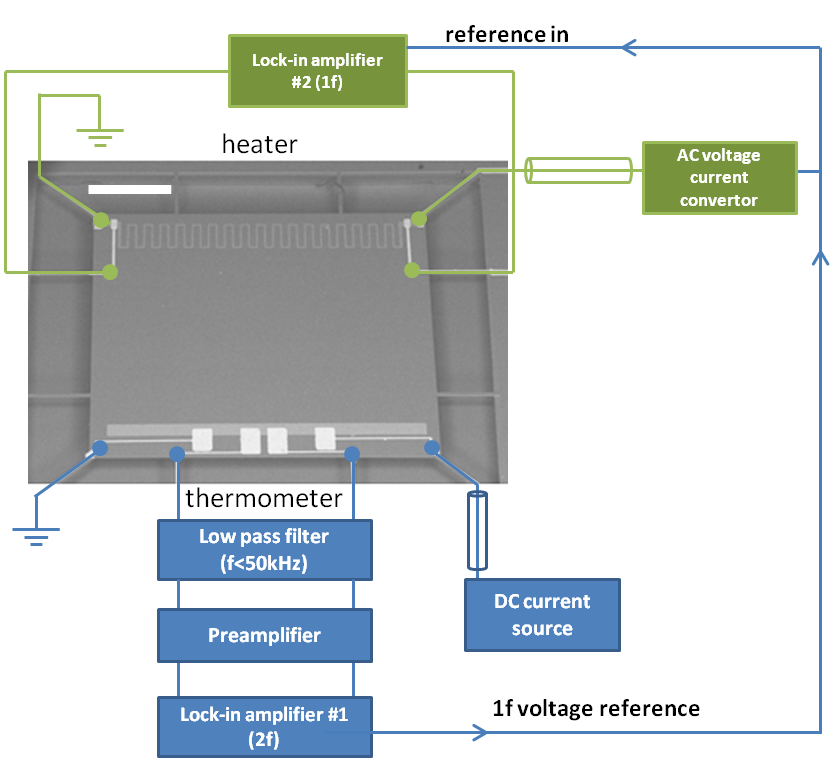}
 \end{center}
 \caption{Schematic drawing of the experimental set-up of heat capacity measurement on a suspended silicon membrane. The internal oscillation signal of the lock-in $\sharp$1 is used to pilot the voltage current convertor used to apply the current to the heater. The lock-in $\sharp$2 is used to measure the resistance of the heater, and then estimate the power dissipated. The oscillation of temperature $T_{ac}$ is measured on the lock-in $\sharp$1 at the second harmonic (2f). The scale bar represents 1~mm.}
 \label{setup}
\end{figure}
The rings have been patterned by electron beam lithography on the thermal sensor made of a suspended silicon membrane (size 4~mm $\times$ 4~mm, thickness $\sim 5$~$\mu$m). The transducers, a copper heater and a highly sensitive NbN thermometer are integrated on each side of the membrane (see Fig.\ \ref{setup}), the silver rings being located between the two elements \cite{bourgRSI}.

\begin{figure}
\begin{center}
\includegraphics[width=\columnwidth]{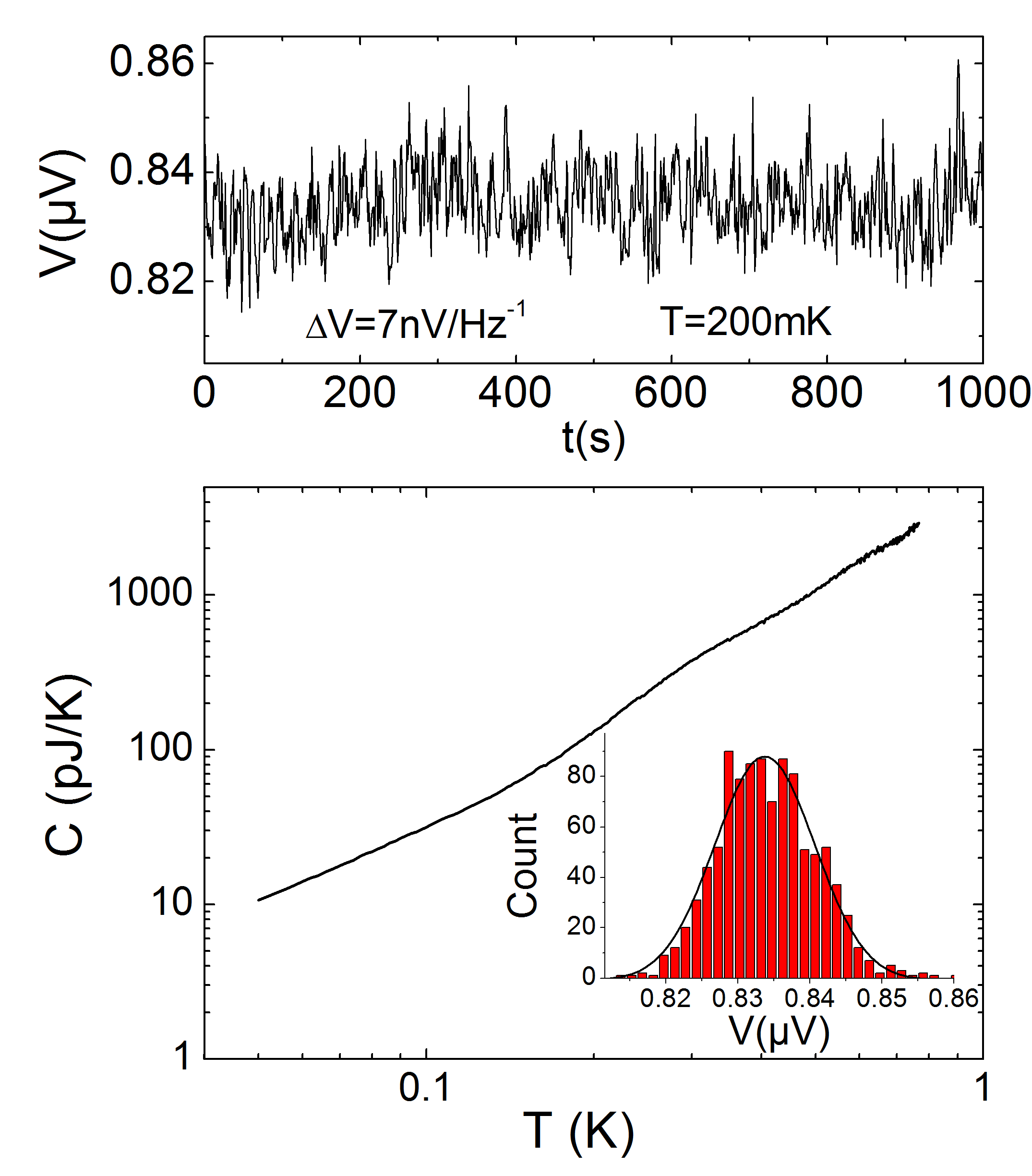}
\end{center}
\caption{(color online) Top panel: voltage at twice the excitation frequency versus time at $T = 200$~mK. Inset of the bottom panel: histogram of the noise extracted from the measurement presented in the top panel. Bottom panel: heat capacity $C$ of silver rings versus temperature in the absence of magnetic field.}
\label{noise}
\end{figure}

A sketch of the ac calorimetry technique that we use is given in Fig.\ \ref{setup} and described in detail in previous articles \cite{seidel,fomin,Garden200916}. It consists in applying an ac current through the heater; here at a frequency $f_{elec}=30$~Hz. This induces oscillations $T_{\rm{ac}}(t)$ of the temperature of the suspended membrane. These oscillations are detected by the thermometer. For a specific experimentally determined operating frequency, quasi-adiabatic conditions are fulfilled, allowing measurements of the specific heat. The typical signal obtained from the lock-in amplifier is composed of a modulus measuring a root mean square (RMS) voltage proportional to the temperature oscillation of the membrane ($T^{\mathrm{RMS}}_{\rm{ac}} \propto  V^{\mathrm{RMS}}_{\rm{ac}}$) and of a phase $\varphi$, both related to the heat capacity through the equations:
\begin{eqnarray}
T_{\rm{ac}}(t) &=& \frac{P_0}{\omega C} \cos(2\omega t+\varphi),
\label{accalo1}
\\
\tan(\Delta \varphi) &=& \frac{2\omega C}{K},
\label{accalo2}
\end{eqnarray}
where $P_0$ is the power dissipated in the heater, $\omega=2\pi f_{elec}$ the electrical excitation frequency, $C$ the heat capacity of the membrane and $K$ the thermal conductance between the membrane and the heat bath (for more details see \cite{seidel,fomin,Bourg05,TCA}).
The setup is cooled down to very low temperatures using a dilution fridge, equipped with a superconducting coil supplying a magnetic field $H$ normal to the plane of the rings. As compared to previous measurements \cite{Bourg05,Ong06,Ong07}, the thermometry has been adapted to work at very low temperature and the arm width has been increased from 40~$\mu$m to 150~$\mu$m to allow measurements at a higher frequency. Consequently, $\delta T_{\rm{dc}} = T_{\rm{membrane}} - T_{\rm{cryostat}}$ decreased and we were able to obtain temperatures compatible with the purpose of our study ($T \simeq 70$~mK on the membrane for a regulated temperature of 50~mK of the sample holder) \cite{TCA}.

\section{Experimental results}

Figure \ref{noise} shows a scan of the heat capacity versus the temperature of the membrane. The heat capacity measured at the lowest temperature is around 10~pJ/K, with the error not exceeding 0.1~pJ/K. It must be noticed that this value corresponds to the heat capacity of the whole sample (rings $+$ membrane). In the inset of Fig.\ \ref{noise}, we show a histogram of the noise measurement obtained from the upper panel of this figure, indicating a noise of $\simeq 0.8$~nV/$\sqrt{\mathrm{Hz}}$.

Our experimental method to detect persistent currents is based on scanning the heat capacity versus the applied magnetic field, at a constant temperature. During the measurement, two components of the temperature oscillation are recorded: the modulus and the phase. Both bring physical informations about the heat capacity variations (see equations (4) and (5)).

\begin{figure}
\begin{center}
 \includegraphics[width=\columnwidth]{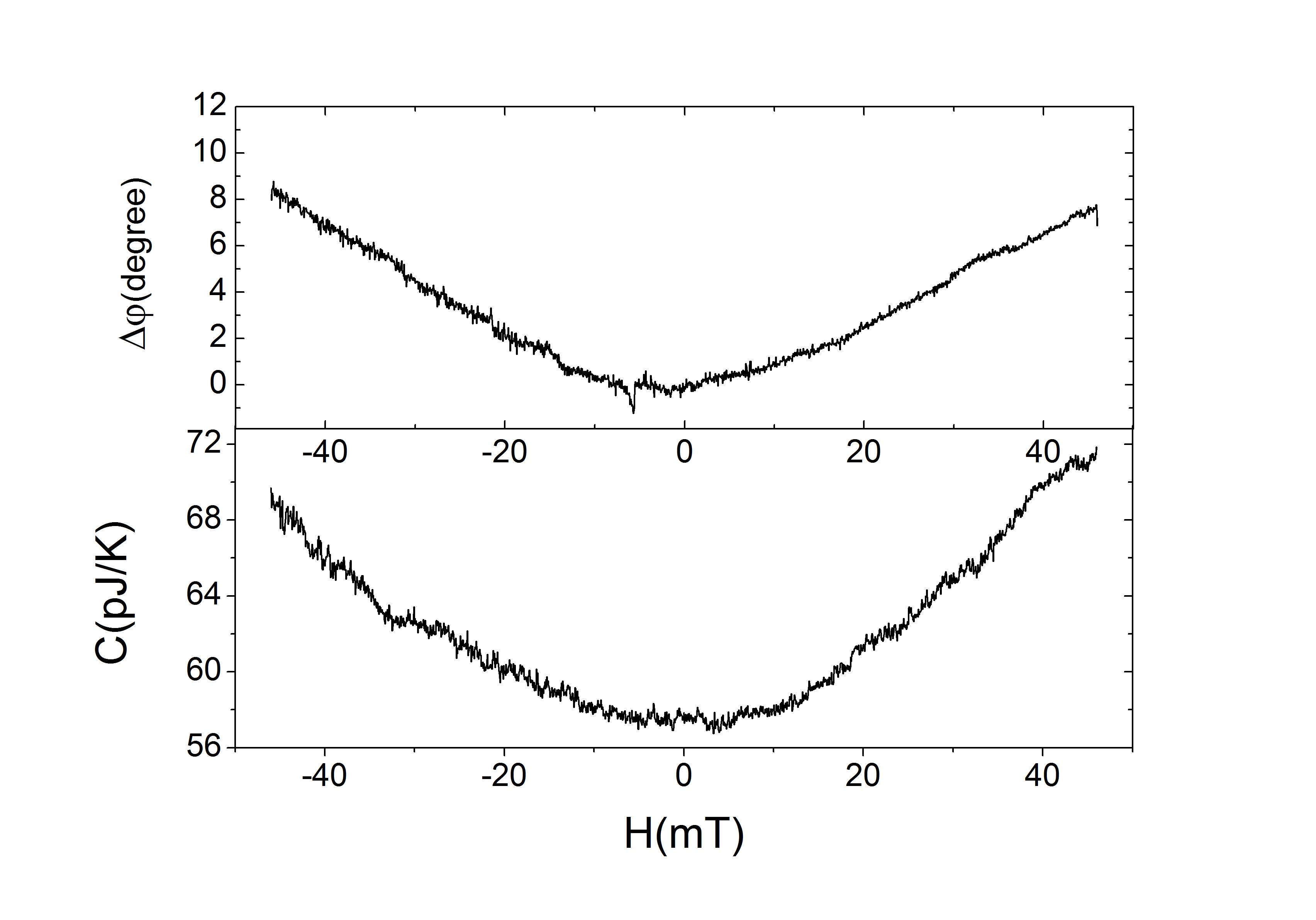}
 \end{center}
 \caption{Phase of the temperature oscillation (at the top) and modulus  of the temperature oscillations, expressed in terms of heat capacity using Eq. (4)(at the bottom), of mesoscopic silver rings versus magnetic field measured at $T = 150$~mK.}
 \label{raw}
\end{figure}

A typical scan is shown in Fig.\ \ref{raw}. Starting at $H = 0$~mT, we slowly increase the field to 50~mT by small steps ($\simeq 1$~mT). $C$ is measured at each $H$ for signal integration times of the order of 10 seconds. In order to improve the sensitivity and to detect the very weak oscillations, we realized a hundred of identical scans by carrying out cycles between $H = -50$~mT and $H =50$~mT. These data are averaged during the signal processing described below.

The measurements were performed at three different temperatures: $T=60$, 100 and 150~mK. For each of them, the temperature of the sensor membrane oscillates with an amplitude of $\sim$ 10\% compared to the control temperature. 

\section{Data processing}

\begin{figure}
\begin{center}
 \includegraphics[width=\columnwidth]{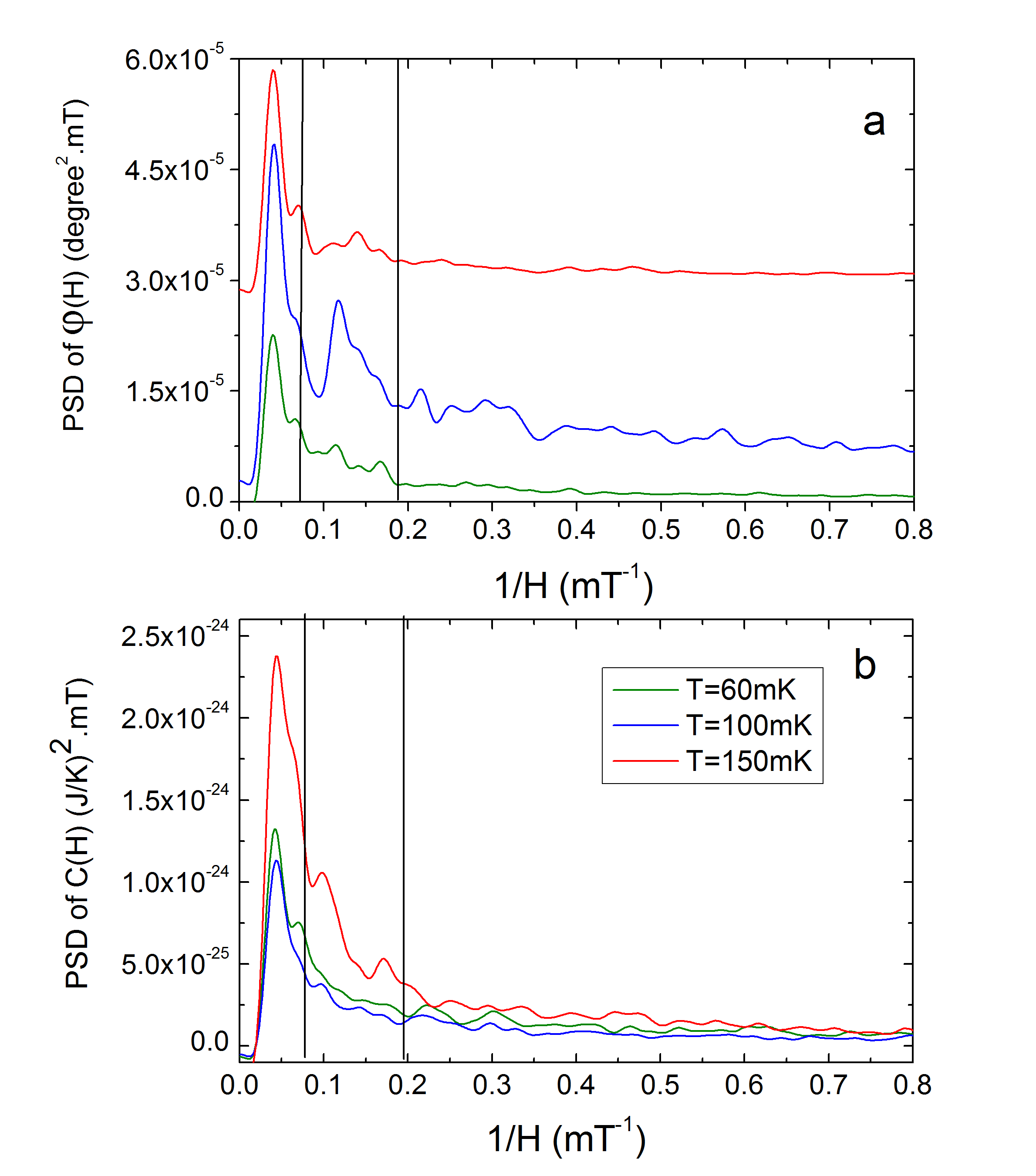}
 \end{center} 
 \caption{(Color online) Power spectral densities of the average signal for phase (upper panel (a)) and in the lower panel (b), the PSD of the modulus of the heat capacity (extracted from the equation 4). In the upper panel the red and the blue curves have been shifted from the green for clarity, and in the medium panel only the red curve has been shifted by $10^{-30}$~(J/K)$^2$.mT. The vertical straight lines delimit the frequency range in which signatures of persistent currents are expected.}
 \label{psd}
\end{figure}

In order to detect a periodicity in our data, a Fourier analysis of the phase of the temperature oscillation and of its amplitude, expressed in terms of heat capacity using Eq. (4), has been performed. As a first step, the low frequency trend observed in every scan has been removed with a third-order polynomial regression. As a second step, the power spectral densities have been calculated as follows. The autocorrelation function of the signal $x(H)$ (where $x$ is the modulus or the phase),
\begin{equation}
\Gamma_{x}(H) = \lim\limits_{M \to \infty} \frac{1}{2M} \int^{M}_{-M} x(H')x(H'-H)dH',
\end{equation}
is first estimated for every scan. Averaging over a hundred of scans is then performed to reduce the noise. The power spectral density $S_x(\nu)$ is calculated as the Fourier transform of the averaged autocorrelation function ${\bar \Gamma}_{x}(H)$.
The power spectral densities (refered to PSD in the following) of both the phase of the temperature oscillation and the modulus of the heat capacity (extracted from equations 4) for the three temperatures $60$, $100$ and $150$~mK are displayed in Fig.\ \ref{psd} .

We first stress that the high peaks observed for all curves at a low frequency $\nu \sim 0.05$ mT$^{-1}$ are a reminiscence of the trend which has not been completely removed by the polynomial regression. These peaks do not carry any useful signal. Unfortunately, they can mask any physically interesting signatures of persistent currents that might be present at low frequencies and would correspond to the first harmonics $C_1$ of the heat capacity.

In addition to the low-frequency peak, our spectral analysis of the phase of the temperature oscillation (upper panel of Fig.\ \ref{psd}) reveals peaks at $\nu \simeq 0.11$~mT$^{-1}$ (the period is 9~mT). This value corresponds to a half of flux quantum through a ring of 264~nm in radius. Our rings have the inner radius of 220~nm and the outer radius of 360~nm. Thus, the observed spectral peak is in the range of frequencies in which signatures of the second harmonics of heat capacity oscillations, $C_2$, are expected.
However, as it can be seen from the lower panel of Fig.\ \ref{psd}, no clear signature of spectral peaks at $\nu \simeq 0.11$~mT$^{-1}$ is observed in the modulus of heat capacity, although the phase and the modulus of the oscillating heat capacity signal are related through Eq.\ (\ref{accalo2}).
The reason for observing a peak in the phase signal but not in the modulus signal most probably resides in a lower sensitivity on the modulus signal. At 100~mK, the sensitivity obtained for the modulus is $\sim 10^{-14}$~J/K or $\sim 10^{-20}$~J/K ($\sim 100$ $k_{\mathrm{B}}$) per ring (a sensitivity of $\Delta C/C \simeq 5\times 10^{-4}$) as compared to the sensitivity obtained for the phase estimated to be of the order of $\Delta \varphi/\varphi \simeq 3\times 10^{-5}$.

\begin{table*}
\begin{tabular}{|c|c|c|c|c|c|c|}
\hline
T (mK) &Area $A$ (degree$^2$) & Phase $\varphi$ (degree) &  $\tan\varphi$ &  $K$(W/K)  &  $|\langle C_2 \rangle|$ (J/K) & $|\langle C_2 \rangle|$/ring ($k_{\mathbf{B}}$) \\
\hline \hline
60 & 2.9$\times10^{-8}$ & 2.4$\times10^{-4}$ & 4.2$\times10^{-6}$& 7.5$\times10^{-9}$&  8.5$\times10^{-17}$ & 1 \\
100 & 25.1$\times10^{-8}$& 7.1$\times10^{-4}$ & 1.2$\times10^{-5}$& 1.8$\times10^{-8}$ & 6.1$\times10^{-16}$ & 9  \\
150   & 4.5$\times10^{-8}$& 3$\times10^{-4}$ &  5.2$\times10^{-6}$& 2.7$\times10^{-8}$ & 3.5$\times10^{-16}$& 5  \\
\hline
\end{tabular}
\caption{The amplitude of the second harmonics of heat capacity oscillations $|\langle C_2 \rangle|$ is extracted from the curves of Fig.\ \ref{psd} (upper panel). First, the area $A$ of the spectral peak is estimated (second column). For this purpose, we fit the curves of the upper panel of Fig.\ \ref{psd} with smooth polynomials and evaluate the integral of the difference between the data and the fit taken between the vertical lines. Next, the associated heat capacity is calculated using Eq.\ (\ref{accalo2}) with $\varphi = \sqrt{2 A}$ (column 5). Finally, the heat capacity is divided by the number of rings $N = 5 \times 10^6$ to obtain the signal per ring (last column).}
\label{table}
\end{table*}

In order to compare our measurements with theoretical predictions, we estimated the average amplitude of the second harmonics of heat capacity oscillations, $|\langle C_2(T) \rangle|$, from the area below the curves of Fig.\ \ref{psd} (see Table \ref{table}).

\section{Comparison with theory}

We compare our results with the predictions of two theoretical models: the model of Ambegaokar and Eckern (AE) \cite{Ambe90} and the model of Bary-Soroker, Entin-Wohlman and Imry (BEI) \cite{Bary08}. Both models rely on taking into account interactions between electrons in the rings, but the interactions are repulsive in the AE model and attractive in the BEI model. In order to suppress the superconductivity that strong attractive interactions may induce, the latter model includes scattering of electrons by magnetic impurities.

\begin{figure}
\begin{center}
\includegraphics[width=\columnwidth]{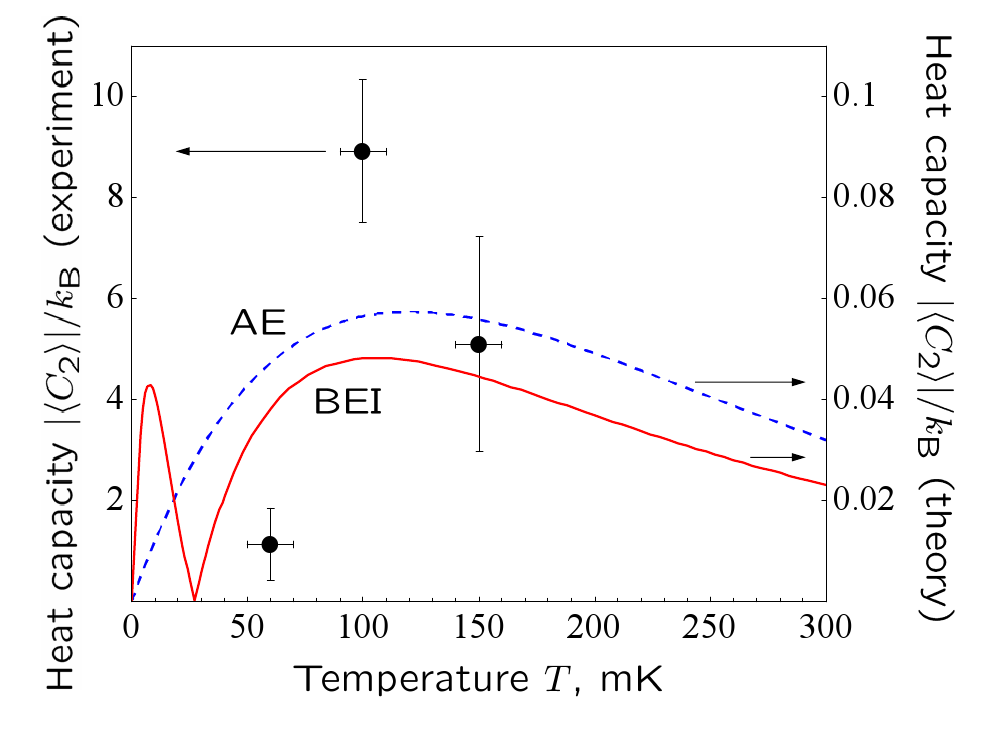}
\end{center}
\caption{(Color online) Average amplitude of the second harmonic $C_2$ of the heat capacity oscillations with the magnetic field. Points show the experimental results from Table \ref{table}, with the scale on the left. Lines show theoretical predictions: Eq.\ (\ref{cae}) (dashed blue line) and Eq.\ (\ref{cbei}) (solid red line), scale on the right.}
\label{theor}
\end{figure}

Even if the applicability of the AE model to realistic experiments is made questionable by the disagreement of the sign of the current that it predicts with measurements \cite{Deblock02}, it is still of great interest to describe experimental data \cite{Bleszynski09}. Using Eq.\ (18) of Ref.\ \cite{Ambe90} and our Eq.\ (\ref{maxwell}), we readily obtain:
\begin{eqnarray}
\langle C_2^{\mathrm{AE}}(T) \rangle = k_{\mathrm{B}} \times \left[ \frac{4}{9 \pi} N(0) {\bar V} \right] \frac{k_{\mathrm{B}} T}{E_{\mathrm{Th}}} \exp\left(- \frac{k_{\mathrm{B}} T}{3 E_{\mathrm{Th}}} \right),
\label{cae}
\end{eqnarray}
where $N(0)$ is the electronic density of states at Fermi energy and ${\bar V}$ is the mean value of the attractive interaction potential \cite{Ambe90}. Using the data from Ref.\ \cite{Deblock02}, where silver rings similar to ours were studied, we estimate $N(0) {\bar V} \simeq 0.37$ and $E_{\mathrm{Th}}/k_{\mathrm{B}} \simeq 40$ mK. The resulting dependence of $\langle C_2^{\mathrm{AE}} \rangle$ on temperature is shown in Fig.\ \ref{theor} by a dashed line, with the scale given on the right. The theoretical Eq.\ (\ref{cae}) reproduces the trend of the temperature dependence of our data that has a maximum at $T \simeq 3 E_{\mathrm{Th}}/k_{\mathrm{B}} \approx 100$ mK. But the values of $|\langle C_2^{\mathrm{AE}} \rangle|$ are 2 orders of magnitude smaller than the  data.

In the BEI model, assuming attractive interactions in combination with scattering on magnetic impurities \cite{Bary08}, we find:
\begin{eqnarray}
&&\langle C_2^{\mathrm{BEI}}(T) \rangle = -k_{\mathrm{B}} \times 4 T \frac{E_{\mathrm{Th}}}{k_{\mathrm{B}}}
\sum\limits_{n=-\infty}^{\infty}
\int_0^{\infty} dx \sin(2 \pi x)
\nonumber \\
&&\times \frac{\partial^2}{\partial T^2}
\left\{
\frac{\Psi'[F_n(x, T, T_c^0, \tau_s)]}{\ln(T/T_c^0) +
\Psi[F_n(x, T, T_c^0, \tau_s)] - \Psi(1/2)}
\right\}, \;\;\;\;\;\;
\label{cbei}
\end{eqnarray}
where $\Psi(x)$ and $\Psi'(x)$ are the digamma function and its derivative, respectively, $\tau_s$ is the spin-scattering time, $T_c^0$ is the bare superconducting transition temperature in the absence of magnetic impurities (\textit{i.e.} at $\tau_s \rightarrow \infty$), and $F_n(x, T, T_c^0, \tau_s) = (1+|n|)/2 + (1/2 \pi \tau_s + \pi x^2 E_{\mathrm{Th}}/k_{\mathrm{B}})/T$. The minus sign in Eq.\ (\ref{cbei}) reflects the diamagnetic nature of the persistent current in this model, in contrast to the paramagnetic current leading to Eq.\ (\ref{cae}). It follows from the comparison of the theory with previous measurements of persistent currents \cite{Levy90,Jariwala01} that it is reasonable to choose $T_c^0 = 0.1 E_{\mathrm{Th}}/k_{\mathrm{B}}$ \cite{Bary08}. At the same time, $s = 1/\pi T_c^0 \tau_s$ should exceed 0.87 to suppress the superconductivity at all temperatures. We set $s = 1$ and show the absolute value of the resulting heat capacity $|\langle C_2^{\mathrm{BEI}} \rangle|$ in Fig.\ \ref{theor} by a solid line.

\section{Discussion}

Interestingly, the two theoretical approaches considered above yield very similar results both for the magnitude of heat capacity oscillations ($|\langle C_2 \rangle| \sim 10^{-2}$ $k_{\mathrm{B}}$) and for its temperature dependence (maximum values of $|\langle C_2 \rangle|$ reached at $T \approx 100$ mK for experimental parameters). Whereas the latter temperature dependence is in agreement with our measurements, the predicted magnitude of the heat capacity oscillations is far too small to explain the observed $|\langle C_2 \rangle| \sim k_{\mathrm{B}}$. This discrepancy may result from the insufficiencies of the theoretical models, as well as from the uncontrolled errors in the estimation of the absolute values of heat capacities in our experiment.

On the theory side, the two models considered here were previously shown to be compatible with experiments (see Refs.\ \cite{Ambe90} and \cite{Bary08}). They provided reasonable results for the magnitude and the temperature dependence of persistent currents measured in Refs.\ \cite{Levy90} and \cite{Jariwala01}. It should be noted, however, that, on the one hand, the heat capacity is proportional to the second derivative of the persistent current $I$ with respect to temperature [see Eq.\ (\ref{maxwell})] and hence, it is sensitive to fine details of the temperature dependence of $I$ that might not be captured by the theory. Equation (\ref{cbei}), for example, changes sign for $T \approx 30$ mK, leading to a peculiar low-temperature behavior of the heat capacity in Fig.\ \ref{theor}. This is due to the change of the shape of $\langle I_2(T) \rangle$ curve from convex to concave. Also the surface states (evanescent states), when included in the theoretical model, may modify its predictions significantly \cite{Chow08}. In addition, it was noticed by several authors that other phenomena (such as, e.g., the fluctuations of electron spin density \cite{Schwab97} or the ambient electromagnetic field \cite{Kravtsov93}) may induce magnetic flux-periodic currents in mesoscopic rings. These currents can be comparable or even larger than the persistent currents. Given the large signals measured in our experiments and taking into account that our calorimetric technique may be particularly sensitive to currents that dissipate heat, we believe that it is likely that these phenomena may be important in our setup. Because Eq.\ (\ref{maxwell}) does not hold for these non-equilibrium processes, it remains to be seen if and how they could be included into the theoretical model.

The disagreement between theory and experiment seen in Fig.\ \ref{theor} might also stem from the difficulties in obtaining quantitatively correct values of heat capacity, intrinsic to the extreme difficulty of the measurements: measuring very low thermal signals at extremely low temperatures. These difficulties are obvious already from the comparison of the heat capacities extracted from the absolute value and the phase of the voltage signal measured in our experiment (compare the two panels of Fig.\ \ref{psd}).

\section{Conclusion}

We introduced a calorimetric approach to the study of persistent currents in mesoscopic rings made of normal metals. The approach relies on the measurement of periodic variations of heat capacity of a large ensemble of rings with magnetic field. Under the experimental conditions reported here, the approach was at the limit of its sensitivity, the signal being strongly masked by noise. Despite this, we estimated the amplitude of heat capacity oscillations to be of the order of several $k_{\mathrm{B}}$ per ring at $\sim 100$ mK. The amplitude is two orders of magnitude larger than expected from the existing theories which, however, correctly predict the range of temperatures where the heat capacity signal is maximum. Further experiments and theoretical investigations are necessary to elucidate the sources of this discrepancy.

Both experiment and theory suggest that, in contrast to the persistent current $I$, the heat capacity $C$ of an ensemble of mesoscopic rings is not a monotonic function of temperature. In particular, the average value of its second harmonic $C_2$ vanishes for $T \to 0$. More experiments having better sensitivity will be necessary to evidence the real position of this maximum. 

\section*{Acknowledgements}

We acknowledge support from Nanofab and Capthercal. We would like to thank P. Brosse-Maron, J.-L. Garden, T. Fournier and U. Gennser for help and fruitful scientific exchanges. We are grateful to G. Montambaux, L. Saminadayar and G. Gaudin for discussions at early stages of this work and to R. Deblock for his comments on the manuscript. This research has been funded by the R\'{e}gion Rh\^one-Alpes (PhD fellowship of GS) and by the ANR through the Quantherm project.


\begin{thebibliography}{99}
\bibitem{footnote}
More precisely, $\langle I_1 \rangle \propto \exp(-L/2 \ell)$, with $\ell$ the elastic mean free path due to disorder \cite{Cheung89,Akker07}. In rings with diffusive scattering, $L \gg \ell$ and $\langle I_1 \rangle = 0$ can be safely assumed.
\end{thebibliography}
\end{document}